# SparkCL: A Unified Programming Framework for Accelerators on Heterogeneous Clusters

Oren Segal, Philip Colangelo, Nasibeh Nasiri, Zhuo Qian, Martin Margala
Department of Electrical and Computer Engineering
University of Massachusetts Lowell
Lowell, MA
oren_segal@student.uml.edu

## ABSTRACT
We introduce SparkCL, an open source unified programming framework based on Java, OpenCL and the Apache Spark framework. The motivation behind this work is to bring unconventional compute cores such as FPGAs/GPUs/APUs/DSPs and future core types into mainstream programming use. The framework allows equal treatment of different computing devices under the Spark framework and introduces the ability to offload computations to acceleration devices. The new framework is seamlessly integrated into the standard Spark framework via a Java-OpenCL device programming layer which is based on Aparapi and a Spark programming layer that includes new kernel function types and modified Spark transformations and actions. The framework allows a single code base to target any type of compute core that supports OpenCL and easy integration of new core types into a Spark cluster.

## 1. INTRODUCTION

It has been shown theoretically [1,2] and increasingly in practice that data centric [3] and HPC [4] applications can benefit from using different types of compute cores in tandem in order to achieve a computational intensive goal. The benefits translate into an increase in the overall power efficiency and performance of a system. Several major examples of such gains can be found in the ever growing list of heterogeneous systems on the green 500 list [5]. A heterogeneous system is typically comprised of conventional cores (CPUs) and unconventional cores (GPUs/FPGAs/DSPs). Such a system employs both conventional and unconventional cores on different types of tasks. Heterogeneous computing brings improved performance and power efficiency through specialization.

One of the main difficulties with hardware specialization is that different types of hardware devices require different types of programming languages, development tools and technical skill sets to master them. Even from a software development stand point alone the increase in the number of lines of code will lead to an increase in system instability [6].

For heterogeneous computing to become a mainstream system design choice, standards need to be adopted for programming various devices. One such standard, OpenCL [7], has started to gain momentum since its introduction in 2008. In the past several years it is becoming an industry choice, with support extended across multiple architectures, albeit not without resistance from manufacturers that supply their own specialized software development environments.

OpenCL although a good step in the right direction is still a difficult challenge for many software engineers [8]. The need to deal with the technical details of OpenCL and the fact that C/C++ is not the language of choice in many development environments such as data centers hampers its chances of success and main stream adoption.

We are now and have been in the past several years in the age of big data processing. The need for processing vast amounts of data and the prediction that we will soon enter the age of exascale computing, a thousand fold increase in the amount of data and calculations on today's standards, is a major motivator in the search for more power efficient computing solutions. As of today we are far from the exascale power efficiency goals set by the DOE for 2020[9]. In order to achieve those goals significant changes need to be made in the way we design and program large scale systems. In addition it is predicted [10] that heterogeneity will play a major role in future exascale systems.

Today's main stream big data processing frameworks such as Spark [11] and Hadoop [12] treat computers as a collection of conventional cores. If they do utilize unconventional cores, it is through specialized libraries typically of the GPU kind [13]. They do not treat unconventional cores as equal citizens in a computing environment and they do not open the possibility of integrating new types of unconventional cores into their midst easily.

SparkCL is an attempt at changing the status quo by allowing a mainstream framework (Spark) to seemingly integrate unconventional cores into its core operations. To achieve that goal we introduce two open source frameworks: aparapi-ucores [17] and spark-ucores (code name SparkCL) [16]. Together they allow a single java code base to be sent to Spark workers and accelerated through specialized hardware if available.

The rest of this paper is organized as follows. Section two gives an overview of the SparkCL framework, section three discusses the execution model and the new functionality added to Spark. Section four discusses some related work, sections five and six discuss conclusions and future directions respectively.

## 2. SPARKCL OVERVIEW

We have combined the power of two separate open source frameworks: Aparapi and Spark to create a new framework that allows accelerating execution using simple java code running on top of Aparapi and the Spark framework.

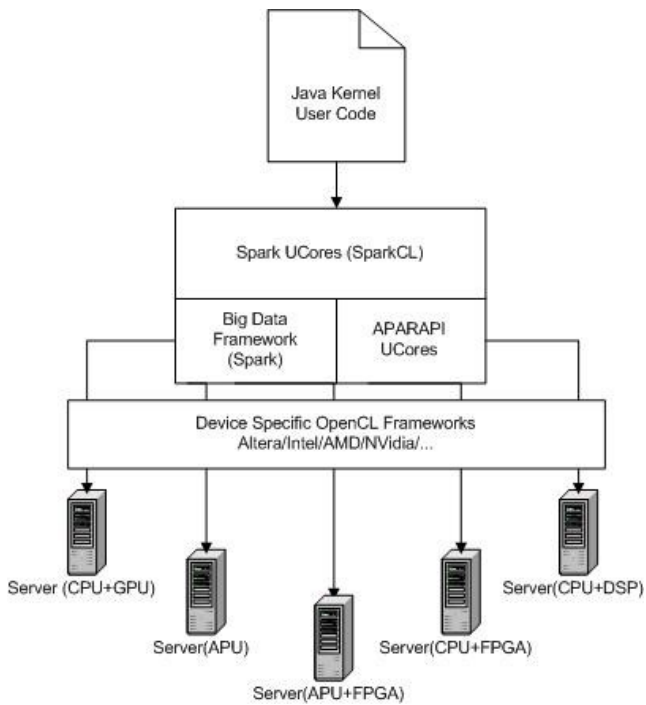

**Fig 1 – Framework overview**

As can be seen in Fig 1, the user execution kernel written in Java gets executed by the SparkCL framework which in turn delegates the code to Spark and eventually to Aparapi UCores on OpenCL capable devices on the cluster. To allow this programming flow, with indifferent treatment of unconventional cores, a new programming layer is built on top of the existing Spark and Aparapi infrastructure.

## 3. SPARK LAYER MODIFICATIONS

Spark, as do other big data frameworks, views computers as collections of cores. It does not take unconventional cores into consideration when it comes to performing general computing functions. To accommodate unconventional cores we had to modify the way functions and data are sent to acceleration devices. Several key issues have been identified in our previous work [14, 15] on accelerating the Hadoop framework:

1) A task has to be computational intensive to justify the overhead of using an accelerator i.e. do to architectural limitations, data transfer overhead, and the way integration of most accelerators into compute fabric is done today, it is difficult to accelerate computationally simple tasks. Instead, the focus should be on highly compute intensive tasks.

2) Enough data must be collected in order to enable efficient acceleration. In most cases a combination of computation complexity and a large amount of data to process is required for efficient acceleration.

3) High level, big data programming frameworks tend to help users generalize their algorithms by distancing them from the concerns of efficient operation of the underlying hardware. That means that things such as memory allocation and device friendly data types are not their strong suite or focus of interest.

To be able to utilize accelerators efficiently, we need to tackle the aforementioned problems before any hope of efficient integration and acceleration can be made.

## 3.1 EXECUTION MODEL AND KERNEL TYPES

In standard Spark a function has a single call site and is replicated across the cluster. This is sufficient for CPUs but does not allow for the special requirements of accelerators, namely large amounts of primitive type data and complex computations. To allow for such needs we have created what we call a ***SparkKernel*** type. This type encapsulates a run function, an Aparapi Kernel (with functionality similar to the existing call override in standard Spark), and two additional functions used to prepare/pre-process the data before running the kernel and post processing of data after the kernel was executed.

```
public abstract class SparkKernel<T,R> extends
SparkKernelBase implements Serializable
{
        public abstract void mapParameters(T data);

        public abstract void run();

        public abstract R mapReturnValue(T data);
}
```

**Figure 2 - Spark Kernel**

An example of an abstract kernel type can be seen in Fig 2. The user overrides the abstract kernel type with his implementations of the three functions.

### 3.1.1 Kernel Function Overrides
The following are the standard functions that should be overridden in the user's kernel.

#### 3.1.1.1 void mapParameters(T data)
Prepare the data and decide which device to use if any. Users can use the API to do things such as set the computation range, choose execution mode (CPU/Accelerator/JTP) and selectively run the kernel if conditions are not ideal to run an accelerated version, for example, in a case where not enough data exists to justify the execution of the kernel.

#### 3.1.1.2 void run()
A standard Aparapi kernel that will be executed on each capable device node on the cluster. The framework will try to cache it by default to avoid multiple instantiation on each worker node.

#### 3.1.1.3 R mapReturnValue(T data)
Post process the data and handle any activity or cleanup needed post kernel execution. If the kernel was not selectively executed, we can call an alternative compute function here.

### 3.1.2 Simple User Kernel Example

```
RddTestArray rddTestArrayRes =
SparkUtil.genSparkCL(rddTest1).reduceCL(new
SparkKernel2<RddTestArray>()
{
        // data
        float[] a;
        float[] b;
        float[] c;

@Override
public void mapParameters(RddTestArray data1, RddTestArray data2)
    {
            a = data1. getDataArray();
            b = data2. getDataArray();
            c = new float[a.length];
            setRange(Range.create(c.length));
    }

    @Override
    public void run()
    {
            int gid = getGlobalId();
            c[gid] = a[gid] + b[gid];
    }

    @Override
public RddTestArray mapReturnValue(RddTestArray data1,
RddTestArray data2)
        {
            return new RddTestArray(c);
        }
});
```

**Figure 3 - Vector Add Example**

Fig 3. Shows a simple vector add kernel, the heart of the kernel is a simple two line code segment that gets executed on accelerated hardware if available. Map parameters and map return facilitate the pre/post processing of the kernel. Complete examples with source can be found accompanied to the open source SparkCL framework distribution [16].

### 3.1.3 SparkCL Transformations and Actions
Standard Spark actions such as Map and transformations such as Reduce needed to be ported to the new framework. SparkCL introduces three such new constructs:

1) MapCL - Map a SparkCL kernel using a Spark style map function.

2) MapCLPartition - Map a SparkCL kernel using a Spark style map partition function.

3) ReduceCL- Map a SparkCL kernel using a Spark style tree-reduce function.

### 3.1.4 Aparapi UCores Integration
Behind the scenes after the Spark-SparkCL integration has been handled by SparkCL, the kernel is sent to a standard Spark worker, together with the Aparapi UCores Java based library. The worker in-turn will execute the Kernel using the Aparapi UCores Framework[17]. Aparapi UCores is a fork of the Aparapi framework [18]. The standard Aparapi library has been expanded to support binary execution flow (needed to support FPGAs and optional for other devices), accelerator types, and platform selection support to allow for easy integration into multiple OpenCL environments and big data frameworks.

### 3.1.5 Worker Setup and Spark Integration
A SparkCL worker is a standard Spark worker that is capable of running both accelerated and non-accelerated tasks i.e. standard Spark tasks. The key to this ability are the following factors:

1) A SparkCL worker is an unmodified Spark Worker class org.apache.spark.deploy.worker.Worker

2) If a standard Spark job is sent to the kernel, it will be executed using the standard worker class functionality.

3) If a SparkCL job is sent to the worker, the SparkCL extensions will take over and handle the acceleration.

4) A SparkCL worker is executed using start-up scripts that make sure it binds to an OpenCL specific/general (ICD capable) implementation on the node.

5) A worker can set its preferred OpenCL orientation on start-up (device/framework) and a user can modify it thorough the kernel code.

To help accomplish the above, included in the distribution is a simple launch script:

*scripts/spark-submit-and-set-env.sh [OpenCL implementation] [Architecture] [Device Type]*

***OpenCL implementation*** – can be *std* or *fpga* for now. Note that std means any ICD compatible OpenCL implementation (AMD/Intel/NVidia etc.)

***Architecture*** - AMD/Intel/NVidia etc. this string gets sent to Aparapi UCores to filter the available platforms. This is where you select the OpenCL platform you want the worker to bind to.

***Device Type*** – selected default acceleration device can be CPU/ACC/JTP. In addition the kernel code can choose to switch between devices programmatically inside the *mapParameters* function.

```
// start an Altera FPGA worker ->
scripts/spark-submit-and-set-env.sh  fpga Altera ACC spark-1.3.0-bin-
hadoop2.4/bin/spark-class   org.apache.spark.deploy.worker.Worker   --
cores 1 spark://192.168.0.112:7077
// start an AMD APU worker (GPU preferred mode) ->
scripts/spark-submit-and-set-env.sh  std AMD GPU spark-1.3.0-bin-
hadoop2.4/bin/spark-class   org.apache.spark.deploy.worker.Worker   --
cores 1 spark://192.168.0.112:7077
// start an AMD APU worker (CPU preferred mode) ->
scripts/spark-submit-and-set-env.sh  std AMD CPU spark-1.3.0-bin-
hadoop2.4/bin/spark-class   org.apache.spark.deploy.worker.Worker   --
cores 1 spark://192.168.0.112:7077
```

**Figure 4 - Worker Execution Modes**

An example of how to start different types of workers on a single node can be seen in Fig 5. The specific node has two OpenCL implementations (Altera/AMD) and three types of devices FPGA, GPU and CPU. We can potentially launch three types of different workers each using a different device type. Note that for the purpose of Spark Scheduling, in the above example, we tell the worker to use one core. This way it will be sent acceleration tasks sequentially and they will not compete on the same hardware acceleration resources. SparkCL has provisions to prevent contention, but this method avoids the potential conflict altogether.

## 3.2 DEMO APPLICATIONS

To demonstrate the operation of the framework we have implemented three algorithms in SparkCL. They can be found in the distributed open source project [16].

### 3.2.1 SparkCLPi

A SparkCL distributed Pi calculation based on the MapCL kernel variants. There are several implementations including an optimized standard Spark one which is significantly faster than the one supplied by the standard Spark distribution on our performance tests. All Pi tests can be launched by name from the command line to allow for easy performance comparison testing.

### 3.2.2 SparkCLVectorAdd

A SparkCL distributed vector addition based on the ReduceCL kernel type. Given sufficient work it executes on the potentially accelerated workers instead of on the single driver (Uses Spark tree reduce instead of reduce).

### 3.2.3 SparkCLWordCount

A SparkCL distributed word count application based on the MapCL kernel type. It is a functional replicate of the Spark demo but implemented in SparkCL. It demonstrates a somewhat more complex kernel with local data and selective execution.

## 4. RELATED WORK

Several high level OpenCL based programming frameworks have been developed in recent years [18, 19, and 20]. Although these libraries support OpenCL their focus is on GPU and CPU development. Several attempts at accelerating the Hadoop framework involved using a modified MapReduce framework to accommodate for the limitations and strengths of accelerators [21, 22]. A previous attempt at accelerating a Hadoop K-Means algorithm on FPGAs showed good promise but involved significant development efforts [14, 15]. Several other attempts were made at creating accelerator frameworks, but all in all these attempts concentrated on GPUs leaving out other types of accelerators hence they did not fully adopt the broader premise of unconventional cores, that each accelerator will have strengths and weaknesses and different types of accelerators should be used to tackle different computation tasks.

## 5. CONCLUSIONS

We introduce SparkCL an open source high level Java-OpenCL framework based on Apache Spark and Aparapi, capable of running high performance and data centric application across different platforms and devices with the hope that it will be helpful to heterogeneous system users and the research community. We offer the potential to use a single high level Java-OpenCL code base across different architectures in a heterogeneous cluster in order to try to maximize code reuse while attempting to harness the power and efficiency of heterogeneous computing.

## 6. FUTURE DIRECTIONS

We plan to continue work on SparkCL framework and hope that others will find it useful and join the effort of developing and extending the SparkCL framework. The project is in its infancy and leaves a lot to be desired in optimization, setup and usability, we envision a framework that would be easy to use and install so that any programmer/researcher concerned with power efficiency would be able to use this system without having intimate knowledge of the underlying architectures and hardware.

## 7. ACKNOWLEDGMENTS
We wish to thank Altera, Nallatech, HP and Terasic for their support and hardware contributions during various stages of this line of research in the past couple of years.

## 8. REFERENCES

[1] Chung, Eric S., et al. "Single-chip heterogeneous computing: Does the future include custom logic, FPGAs, and GPGPUs?." Proceedings of the 2010 43rd Annual IEEE/ACM International Symposium on Microarchitecture. IEEE Computer Society, 2010.

[2] S. Huang, S. Xiao, and W. Feng. 2009. On the energy efficiency of graphics processing units for scientific computing. In Proceedings of the 2009 IEEE International Symposium on Parallel&Distributed Processing (IPDPS '09). IEEE Computer Society, Washington, DC, USA, 1-8.

[3] Chalamalasetti, S.; Margala, M.; Vanderbauwhede, W.; Wright, M.; Ranganathan, P., "Evaluating FPGA-acceleration for real-time unstructured search," *Performance Analysis of Systems and Software (ISPASS), 2012 IEEE International Symposium on*, vol., no., pp.200,209, 1-3 April 2012.

[4] Gan, Lin, et al. "Accelerating solvers for global atmospheric equations through mixed-precision data flow engine." Field Programmable Logic and Applications (FPL), 2013 23rd International Conference on. IEEE, 2013.

[5] The Green500 List (11/2014). http://www.green500.org/news/green500-list-november-2014

[6] Steve McConnell. 1993. Code Complete: A Practical Handbook of Software Construction. Microsoft Press, Redmond, WA, USA.

[7] Khronos OpenCL Working Group. "OpenCL-The open standard for parallel programming of heterogeneous systems." On line] http://www. khronos. org/opencl (2011).

[8] TIOBE Index. http://www.tiobe.com/index.php/content/paperinfo/tpci/index.html

[9] J. a. D. S. a. M. J. Shalf, "Exascale computing technology challenges," in High Performance Computing for Computational Science--VECPAR 2010, Springer, 2011, pp. 1--25.

[10] Dongarra, Jack. "The international exascale software project roadmap." International Journal of High Performance Computing Applications (2011): 1094342010391989.

[11] Zaharia, Matei, et al. "Spark: cluster computing with working sets." Proceedings of the 2nd USENIX conference on Hot topics in cloud computing. 2010.

[12] Apache Hadoop. http://hadoop.apache.org/

[13] Canny, John. "Interactive Machine Learning." University of California, Berkeley (2014).

[14] Segal, Oren, et al. "High level programming framework for FPGAs in the data center." Field Programmable Logic and Applications (FPL), 2014 24th International Conference on. IEEE, 2014.

[15] Segal, Oren, et al. "High Level Programming for Heterogeneous Architectures." arXiv preprint arXiv:1408.4964 (2014).

[16] Spark for Unconventional Cores (SparkCL). https://gitlab.com/mora/spark-ucores

[17] Aparapi for Unconventional Cores. https://gitlab.com/mora/aparapi-ucores

[18] Aparapi. API for data parallel Java, http://code.google.com/p/aparapi/.

[19] JavaCL. http://code.google.com/p/javacl/

[20] Ruby-OpenCL. http://ruby-opencl.rubyforge.org/

[21] He, Bingsheng, et al. "Mars: a MapReduce framework on graphics processors." Proceedings of the 17th international conference on Parallel architectures and compilation techniques. ACM, 2008.

[22] Basaran, Can, and Kyoung-Don Kang. "Grex: An efficient MapReduce framework for graphics processing units." Journal of Parallel and Distributed Computing 73.4 (2013): 522-533.